\documentclass[aps,pra,showpacs,twocolumn]{revtex4}
\usepackage{graphicx}
\usepackage{bm}
\usepackage{amsfonts}

\usepackage{amssymb,amsthm,dsfont,bm,color,ulem}

\begin{document}

\title{Majorization of quantum polarization distributions}

\author{Alfredo Luis and Gonzalo Donoso}
\affiliation{Departamento de \'{O}ptica, Facultad de Ciencias
F\'{\i}sicas, Universidad Complutense, 28040 Madrid, Spain }

\begin{abstract}
Majorization provides a rather powerful partial-order  classification of probability distributions depending only
on the spread of the statistics, and not on the actual numerical values of the variable being described. We 
propose to apply majorization as a meta-measure of quantum polarization fluctuations, this is to say of the 
degree of polarization. We compare the polarization fluctuations of the most relevant classes of quantum and 
classical-like states. In particular we  test  Lieb's conjecture regarding classical-like states as the most 
polarized and a complementary conjecture that the most unpolarized pure states are the most nonclassical. 
\end{abstract}


\maketitle

\section{Introduction}

Light fluctuations are relevant both from fundamental as well as practical perspectives. On the one hand 
field statistics are the key feature distinguishing classical from quantum light \cite{MW}. On the other hand, 
fluctuations and uncertainty usually limit the performance of  optical applications. In this regard is worth 
noting that polarization and two-beam linear interferometry share the same fundamental SU(2) symmetry, so 
we may say that they are isomorphic. Deep down, this equivalence holds because interference and polarization 
are the two main manifestations of coherence. 

Both in classical and quantum optics polarization uncertainty is assessed via the degree of polarization 
\cite{PG,BR,PiO}.  The classic definition in terms of the Stokes parameters involves just second-order statistics
of the field complex amplitudes. This cannot reflect statistical properties involving higher-order moments, 
in particular polarization fluctuations, which are crucial in quantum optics \cite{PiO}. For example there are 
states with vanishing degree of polarization that nevertheless cannot be regarded as being unpolarized, 
which is usually referred to as hidden polarization \cite{PiO,hp}.

These and similar reasonings have motivated the introduction of other measures of polarization fluctuations, 
actually plenty of them \cite{BR,PiO,YQ,odp}.  In this work we go beyond  particular definitions of the 
degree of polarization by applying the mathematical idea of majorization to  quantum polarization distributions. 
Majorization provides a rather powerful partial-order  classification of probability distributions depending only
on the spread of the statistics, and not on the actual numerical values of the variable being described 
\cite{mj}. This ordering is respected by the entropic measures. So majorization actually becomes 
a kind of meta-measure of uncertainty. In our case this means to go beyond all measures of the degree of polarization 
introduced so far.

As a suitable polarization distribution in quantum optics we focus on the SU(2) $Q$ function because of its 
good properties, specially SU(2) invariance \cite{YQ,Q}. We apply this technique to the most relevant 
classical and nonclassical polarization states. In particular we test  Lieb's conjecture regarding SU(2) 
coherent states as the most polarized states in quantum optics \cite{Lc}. Since SU(2) coherent states are 
also regarded as the most classical states \cite{cs,csmc}, this suggests the ensuing complementary conjecture: 
that the most quantum states should  be the most unpolarized pure states \cite{upnc}. This  conjecture can 
readily be tested also via majorization. 

In Sec. II we present the main ingredients such as  the polarization SU(2) $Q$ function, majorization, and the 
most relevant classes of states to be compared. This includes the SU(2) coherent states as the most 
classical-like, as well as non-classical examples such as squeezed states, the so-called $N00N$ states, 
the phase states, and finally the most  non-classical states according to the Hilbert-Schmidt distance. In 
Sec. III the polarization distributions of these states  are compared via majorization. Since in principle 
polarization and intensity are independent degrees of freedom, we focus mainly  on states with definite 
total number of photons. Nevertheless, we consider also more practical and experimentally generable states 
with non definite total number of photons.

\section{Procedure}

\subsection{Polarization distribution}

A suitable polarization distribution can be introduced via the SU(2) $Q$ function $Q (\Omega )$  defined by 
projection of the density matrix $\rho$ on the SU(2) coherent states as   \cite{YQ,Q}
\begin{equation}
\label{Q}
Q(\Omega)=\sum_{n=0}^\infty \frac{n+1}{4\pi} \langle n,\Omega |\rho | n,\Omega \rangle ,
\end{equation}
where $|n,\Omega \rangle$ are the SU(2) coherent states \cite{cs}
\begin{eqnarray}
| n, \Omega \rangle & = & \sum_{m=0}^n \left( \begin{array}{c} n \\ m \end{array}\right )^{1/2}
 \left ( \sin  \frac{\theta}{2} \right )^{n-m} \left ( \cos  \frac{\theta}{2} \right )^m   \nonumber  \\
& &  e^{-i m \phi} |m , n-m \rangle,
\end{eqnarray}
and $ | n_1 , n_2 \rangle = | n_1 \rangle_1 | n_2 \rangle_2$ denote the product of photon number states in the 
corresponding  two field modes sustaining the polarization degree of freedom. The variables  $\Omega = (\theta, 
\phi)$ represent points on an unit sphere, the Poincar\'e sphere, with polar angle $\theta$, azimuthal angle $\phi$, 
and surface element $d \Omega = \sin \theta d \theta d \phi$. 

The SU(2) symmetry reflects the fact that all points on the sphere are equivalent. This is conveniently respected by 
the SU(2) $Q$ function since the $Q (\Omega)$ function for the transformed state has the same form of the 
original one, but simply centered at another point of the Poincar\'e sphere.  SU(2) transformations are quite simply 
implemented in practice via phase plates or beam splitters.

To simplify the comparison between distributions via majorization  we shall discretize the polarization distribution 
by dividing the Poincar\'e sphere into $N$ surface elements, say pixels. The key point to maintain the natural 
SU(2) invariance is that all pixels should be of the  same area. Taking into account that 
$d \Omega = \sin \theta d \theta d \phi = | d \cos \theta | d \phi$, we accomplish this by dividing the ranges of variation 
of $\cos \theta$ and $\phi$ into intervals of the same length. This is 
\begin{eqnarray}
\theta_\ell & = & \arccos \left ( \frac{2\ell - 1}{N_\theta}  - 1 \right ), \quad \ell = 1, \ldots, N_\theta,  \nonumber \\
\phi_k & = & \frac{2\pi}{N_\phi} k - \pi, \quad k = 1, \ldots , N_\phi .
\end{eqnarray}
Thus the discretized version of $Q ( \Omega)$ is 
\begin{eqnarray}
 p_j & = & Q \left (\Omega_j \right ) d \Omega ,  \quad \Omega_j = (\theta_\ell, \phi_k) \nonumber \\
 j & = & N_\phi \left ( \ell - 1 \right ) + k = 1, \ldots, N , 
\end{eqnarray}
where  $N = N_\theta N_\phi$ and $d \Omega = 4 \pi/N$. More rigorously we should integrate the $Q(\Omega)$ 
distribution to each pixel, but this approximate form is rather simple and good enough for our purposes if the 
sampling is accurate. In the limit of accurate sampling neither the area nor the shape of the pixels matter.
 
\subsection{Majorization}

\begin{figure}
\centering
\includegraphics[width=6cm]{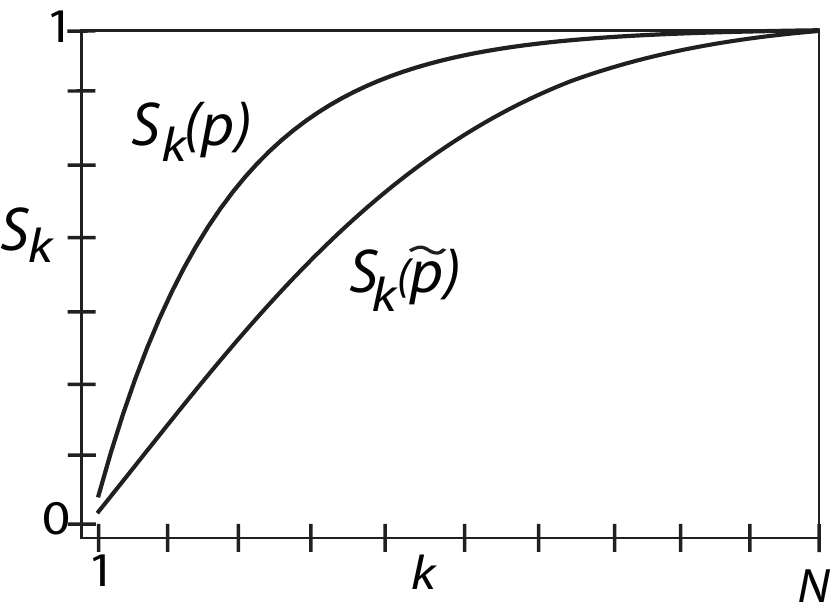}
\caption{Relation between partial ordered sums $S_k$ as functions of $k$ when the majorization  $\tilde{p} 
\prec p$ holds. Although $k$ is discrete, in all plots the points have been joined by continuous lines as 
an aid to the eye.}
\end{figure}

\begin{figure}
\begin{center}
\includegraphics[width=6cm]{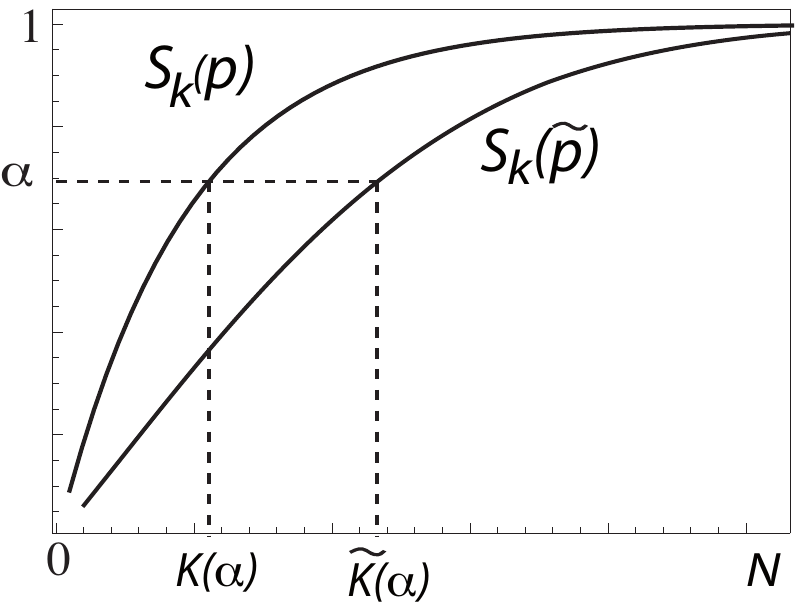}
\caption{Relation between partial ordered sums $S_k$ and confidence intervals 
when the majorization  $\tilde{p} \prec p$ holds.}
\end{center}
\end{figure}

Since polarization lives on an sphere, this is a good place to apply  statistical evaluations of uncertainty and 
fluctuations beyond variance and standard first-order moments. For example confidence intervals or entropy-like 
measures.  In this regard, both lead us to majorization as a kind of meta-measure of fluctuations. Let us show this 
in more detail: we first present two equivalent formal definitions of majorization and then we provide some physical 
intuition about it.
 
Denoting by  $\tilde{p}$ and $p$ two given probability distributions, we say  that $p$ majorizes  $\tilde{p}$, which is 
expressed as $\tilde{p} \prec p$, when the following relation between the ordered partial sums, or  Lorenz curves,  
is satisfied for all $k$ (see Fig. 1):
\begin{equation}
\label{pos}
S_k \left(\tilde{p}^{\downarrow}\right)=\sum_{j=1}^k \tilde{p_j}^{\downarrow}\leq \sum_{j=1}^{k}
p_j^{\downarrow}=S_k\left(  p^{\downarrow}\right )  ,
\end{equation}
where $k= 1,\, 2, \ldots,N$ represents the number of pixels the probabilities of which are added in the corresponding ordered 
partial sum $S_k$, always with $S_N=1$. The superscript $\downarrow$ denotes that the $p_j$ values are arranged 
in decreasing order: $p^\downarrow_1 \geq p^\downarrow_2 \geq \dots \geq p^\downarrow_N$.  We will say that two 
distributions are comparable if one majorizes the other. Moreover,  $\tilde{p}\prec p$ is equivalent to say that there 
exist $N$-dimensional permutation matrices  $\Pi_j$ and a probability distribution $\{\pi_j\}$ such that
\begin{equation}
\label{per}
\tilde{p}=\sum_j \pi_j \Pi_j p .
\end{equation}
That is, $\tilde{p}$ is majorized by $p$ when $\tilde{p}$ can be obtained from $p$ by randomly permuting 
its components, and then averaging over the permutations. 

Majorization is a partial ordering relation, so that not every two distributions can be compared. Thus  
we can find distributions that neither $\tilde{p} \prec p$ nor $p \prec \tilde{p}$. This situation will be 
represented as  $p \Join \tilde{p}$. In such a case the Lorenz curves $S_k$ will intersect. 

Roughly speaking, if  $p$ majorizes $\tilde{p}$ we may say that $p$ presents less dispersion or 
less uncertainty than $\tilde{p}$ regarding the underlying physical property. This is because the partial 
sums (\ref{pos}) indicate that more probability is concentrated in a lesser number of pixels. This idea that 
$\tilde{p}$ is more random that $p$ is also clearly expressed  by the randomization procedure in Eq.(\ref{per}). 

This interpretation can be further illustrated if we consider the two extremes situations. If there were no 
uncertainty, all the distribution should be concentrated in a single pixel, $p^\downarrow_1 = 1$, 
$p^\downarrow_{j \neq 1} =0$, and $S_k = 1$ for all $k$. This clearly majorizes any other distribution. 
On the other hand, the uniform distribution $p^\downarrow_j = 1/N$ is majorized by any other distribution 
\cite{BZ06}. 

This intuition is further confirmed by the deep relation between majorization and other measures 
of uncertainty. Let us present two clear examples: confidence intervals $K (\alpha)$ and entropies $R_q (p)$. 

Confidence intervals $K (\alpha)$ are defined as the minimum number of pixels $K$ such that  the partial 
sum up to $p^\downarrow_K$ comprises a given fraction $\alpha$ of the probability  \cite{LP}. This is:
\begin{equation}
 \sum_{j=1}^K  p^{\downarrow} _j \geq \alpha \longleftrightarrow  K \geq K (\alpha) .
 \end{equation}
When two distributions are comparable, $\tilde{p} \prec p$ is equivalent to saying that all confidence 
intervals of $\tilde{p}$ are larger  than or equal to  those of $p$ (see Fig. 2):
\begin{equation}
\label{cond}
\tilde{p} \prec p  \longleftrightarrow \tilde{K}  (\alpha) \geq K (\alpha) \; \;  \forall \alpha .
\end{equation}
Otherwise, if the distributions are incomparable $p \Join \tilde{p}$ we will have $ \tilde{K} (\alpha) >  K (\alpha)$ and  
$\tilde{K}  (\beta) < K  (\beta)$ for different $\alpha$ and $\beta$.

Regarding the relation between entropy-like measures and majorization we may consider for example the  R\'{e}nyi   
entropies \cite{RB}
\begin{equation}
\label{R}
R_q (p)= \frac{1}{1-q} \ln \left ( \sum_{j=1}^N p_j^q\right ),
\end{equation}
where $q > 0$ is an index labeling different entropies, so we have that if  $\tilde{p}  \prec p$ then $R_q (\tilde{p} ) 
>  R_q (p )$ for all $q$. The limiting case $q \rightarrow 1$  is the Shannon entropy $R_1 =  -  \sum_{j=1}^N  p_j 
\ln p_j$ while $q=2$ is essentially the degree of polarization introduced in Ref. \cite{YQ}.  If the distributions are 
incomparable $p \Join \tilde{p}$  different entropies may provide contradictory conclusions: $R_q (\tilde{p} )  > 
R_q (p) $ while  $R_r (\tilde{p} )  < R_r  (p) $ for some $r \neq q$. 

We think this reveal the powerfulness of majorization as a kind of meta-measure. When majorization holds 
there is unanimity of  confidence intervals and entropies regarding which distribution is more ordered and has 
lesser uncertainty. When there is no majorization the unanimity is lost.

\subsection{Distributions for relevant field states}

Let us recall the classes of classical-like and non-classical field states the polarization distributions 
of which will be compared. We will focus mainly on field states defined within the subspaces ${\cal H}_n$ of 
fixed total photon number $n$. These subspaces have dimension $n+1$ being spanned by the product 
of number states $|m, n-m \rangle$, $m=0,\ldots, n$. We consider pure states  to focus 
exclusively on uncertainty with quantum origin.

\subsubsection{SU(2) coherent states}

These are considered  as the most classical polarization states \cite{csmc}. According to  Lieb's conjecture 
they should majorize any other one within ${\cal H}_n$ \cite{csmc,Lc}. After Eq. (\ref{per}) this is particularly 
clear for classical-like states of the form $\rho = \int d \Omega P (\Omega ) | n, \Omega \rangle \langle n, 
\Omega |$, with a {\it bona fide} classical probability distribution $P(\Omega)$. This is because all the SU(2) 
coherent states are connected by an SU(2) transformation, so the corresponding discretized $Q (\Omega)$ 
are just connected by pixel permutations. Maybe, the surprising result is that this extends to nonclassical light 
with highly singular $P (\Omega )$ distributions. 

Using the SU(2) symmetry we will consider the $Q$ function for the SU(2) coherent state $C$  which is just 
the product of a number state with $n$ photons and the vacuum state: 
\begin{equation}
|n, C \rangle  = | n,0 \rangle .
\end{equation}
The corresponding $Q$ function is concentrated at the north pole of the Poincar\'e sphere. The other SU(2) 
coherent states $|n,\Omega \rangle$ are just SU(2) orbits of this state.

\subsubsection{Phase states}

These are complementary to the number states  \cite{ps}:
\begin{equation}
|n, \phi \rangle = \frac{1}{\sqrt{n+1}} \sum_{m=0}^n e^{- i m \phi} |m,n-m \rangle .
\end{equation}
Using the SU(2) symmetry we will consider the $Q$ function for the phase state $P$ with $\phi=0$, this is $|n,P \rangle = 
| n, \phi=0 \rangle$. 

\subsubsection{Squeezed states}

These are quite distinguished states regarding quantum applications, including metrology as a relevant example. 
There are no simple criteria translating the simple quadrature squeezing into SU(2) squeezing
\cite{LK}. For our purposes we can focus on the most squeezed states $S$ regarding metrological applications, 
exemplified by the twin-number states 
\begin{equation}
|n, S \rangle = |n/2 , n/2 \rangle 
\end{equation}
for even $n$ \cite{HB},  and the closets analog for $n$ odd \cite{JD} 
\begin{equation}  
|n, S  \rangle = \frac{1}{\sqrt{2}} \left  ( \left  | \frac{n+1}{2} ,  \frac{n-1}{2} \right \rangle  + 
\left  | \frac{n-1}{2}  , \frac{n+1}{2} \right  \rangle \right ) .
\end{equation}

\subsubsection{$N00N$ states}

Further states with interesting practical applications relying on their strong quantum properties are the $N00N$ or 
Schr\"odinger's cat states \cite{N00N} which we shall refer to as $N$
\begin{equation}  
|n,  N \rangle = \frac{1}{\sqrt{2}}\left( |n , 0 \rangle + | 0 , n \rangle \right ) .
\end{equation}

\subsubsection{Most non-classical states via Hilbert-Schmidt distance}

 These are the most nonclassical states according to the Hilbert-Schmidt distance to the convex set of 
 classical-like states defined as the incoherent mixture of SU(2) coherent states \cite{QQ}. They have no 
 simple general expression and we will consider just the examples with lower number of photons, say 
 \begin{equation}
 | n= 4,  H \rangle = \frac{1}{\sqrt{3}} \left ( | 0, 4 \rangle + \sqrt{2} | 3, 1 \rangle \right ),
 \end{equation}
 and 
\begin{equation}
 | n= 5,  H \rangle = \frac{1}{\sqrt{2}} \left ( | 1, 4 \rangle +  | 4 , 1 \rangle \right ),
 \end{equation}
while for $n=2,3$ they coincide with the $N00N$ states. It is worth noting that these states coincide with 
the so-called anti-coherent states, defined as those with mean value and variance of  Stokes-operators 
vector invariant  under SU(2) transformations, and some other approaches \cite{upnc}.

\section{Results}

In this section we present the results obtained for the lowest photon numbers, that nevertheless clearly illustrate 
the situation regarding the mutual relationship between classical-like and nonclassical states.  
 
\subsection{One-photon states $n=1$}

The case of a single photon is trivial since all pure states are SU(2) coherent states, so all pure states 
have the same polarization distribution, modulus SU(2) transformations.  

\subsection{Two-photon states $n=2$}

In this case after SU(2) symmetry all the above classes of states reduce to the comparison  of the SU(2) coherent 
state $| 2, C \rangle = |2,0 \rangle$, the phase state $|2,P \rangle$, and the product of one-photon states $|1,1 
\rangle$ that is simultaneously $N00N$, squeezed, and the most non-classical state $| 2,N \rangle = | 2, S \rangle 
= | 2, H \rangle = | 1,1 \rangle$. Their  ordered partial sums $S_k$ are plotted in Fig. 3 as functions of $k$, where 
it can be appreciated that the following sequence holds 
\begin{equation}
N=S=H \prec P \prec  C ,
\end{equation} 
so that the most polarized is the most classical and the most unpolarized is the most non-classical. Note also 
that coherent and phase states are so close that they can be hardly distinguished.   

\begin{figure}
\centering
\includegraphics[width=7 cm]{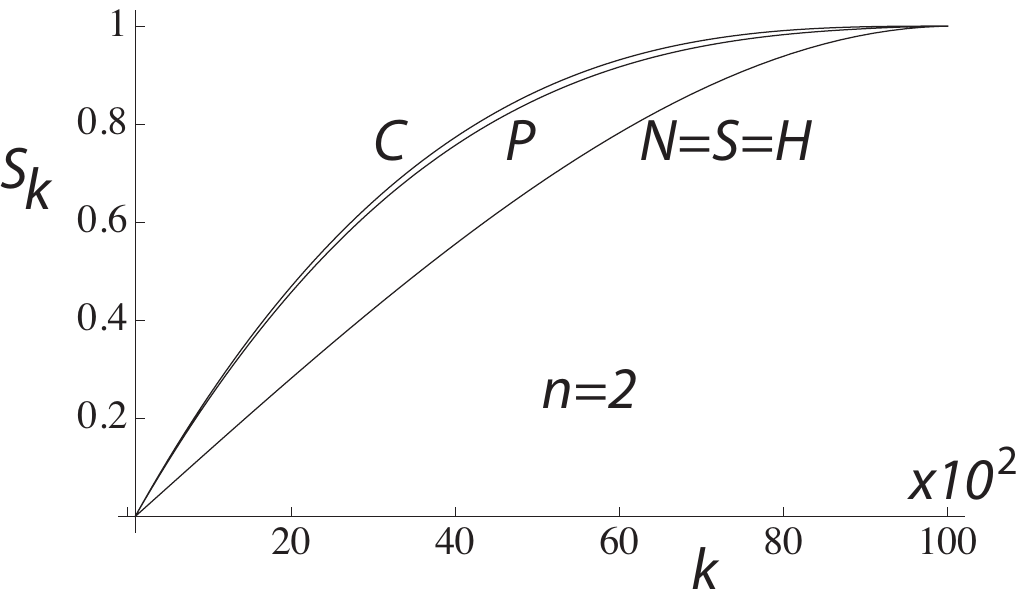}
\caption{Ordered partial sums $S_k$  as functions of $k$ for coherent $C$, squeezed $S$, $N00N$ $N$, most 
quantum $H$, and phase states $P$ for  two-photon states $n=2$.}
\end{figure}

\subsection{Three-photon states $n=3$}

In this case the identity between nonclassical-states holds only between the $N00N$ and the most non-classical. 
Their  ordered partial sums $S_k$ are plotted in Fig. 4 showing the following chain of majorizations
 \begin{equation}
N = H \prec S \prec P \prec  C .
\end{equation}

\begin{figure}
\centering
\includegraphics[width=7 cm]{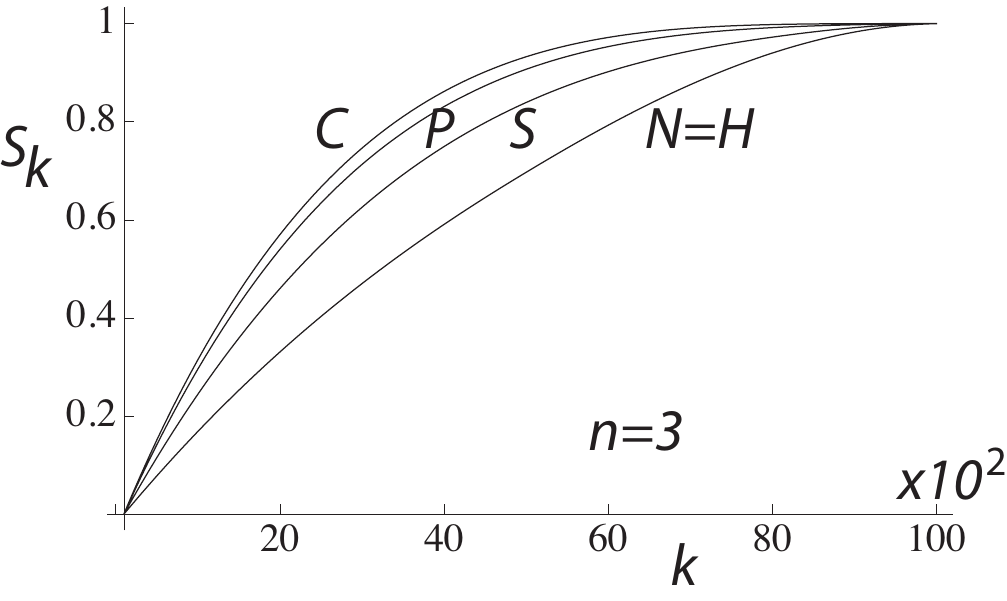}
\caption{Ordered partial sums $S_k$  as functions of $k$ for coherent $C$, squeezed $S$, $N00N$ and most quantum 
$N=H$ and phase states $P$ for  three-photon states $n=3$.}
\end{figure}

\subsection{Four-photon states $n=4$}

In this case all the above classes of states are represented by different vectors. Their  ordered partial sums $S_k$ are 
plotted in Fig. 5 showing the following ordering
\begin{equation}
H \prec S \Join N \prec P \prec  C .
\end{equation} 
We get the first example of incomparability, that holds between the $N00N$ $N$ and squeezed $S$ states. We have 
checked that exactly the same situation is repeated for six-photon states $n=6$.

\begin{figure}
\centering
\includegraphics[width=7 cm]{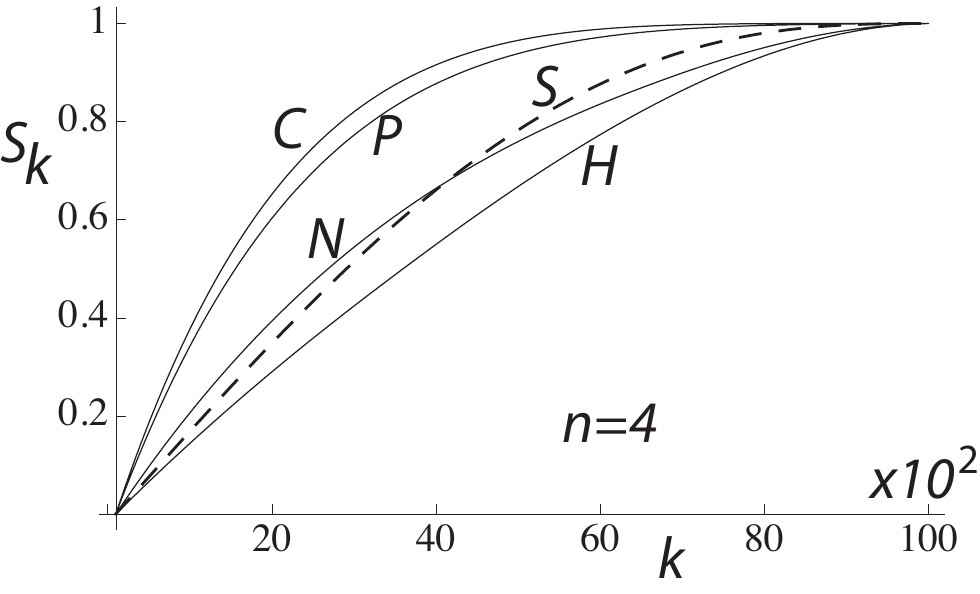}
\caption{Ordered partial sums $S_k$  as functions of $k$ for coherent $C$, squeezed $S$,  $N00N$ $N$, most quantum 
$H$ and phase states $P$ for  four-photon states $n=4$. For clarity the squeezed case $S$ is plotted with dashed line.}
\end{figure}

\subsection{Five-photon states $n=5$}
For the cases we have examined with odd $n$ there is no incomparability between squeezed and $N00N$ states. For $n=5$ 
we have the chain of majorizations 
\begin{equation}
H \prec N \prec S \prec P \prec  C ,
\end{equation} 
as illustrated in Fig. 6. 

\begin{figure}
\centering
\includegraphics[width=7 cm]{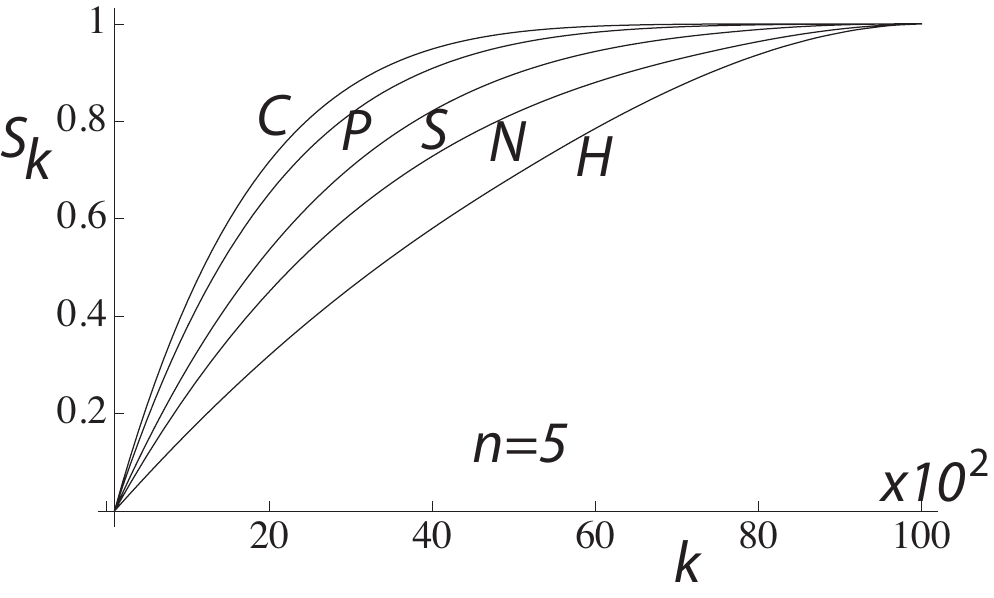}
\caption{Ordered partial sums $S_k$  as functions of $k$ for coherent $C$, squeezed $S$,  $N00N$ $N$, most quantum 
$H$ and phase states $P$ for  five-photon states $n=5$.}
\end{figure}

\subsection{Inter-photon-number}

For states of the same class we have observed the natural behavior that states with larger photon numbers 
majorize states with lower numbers. In this regard we consider the squeezed states with even and odd $n$ 
as different classes. Naturally, the situation is richer when comparing states of different classes and
different photon numbers,  so that incomparability may appear. A simple example is provided in Fig. 7 showing 
incomparability between a coherent state with $n=2$ and a $N00N$ state with $n=6$.

\begin{figure}
\centering
\includegraphics[width=7 cm]{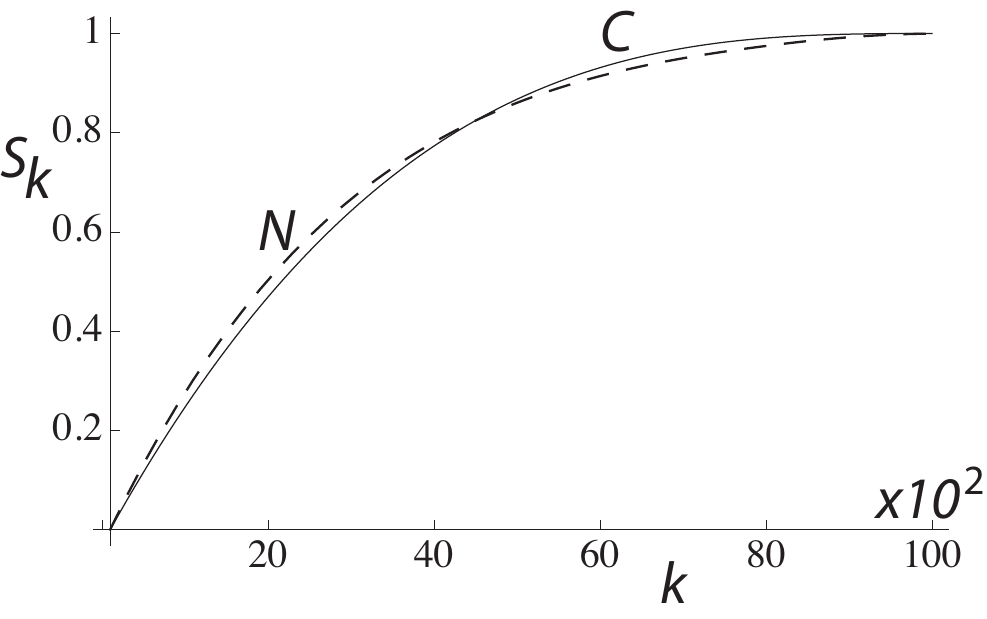}
\caption{Ordered partial sums $S_k$  as functions of $n$ for coherent state with $n=2$ photons $C$ and a $N00N$ state 
$N$ with $n=6$ photons showing that they are incomparable although very similar. For clarity the $N00N$ case is plotted 
with dashed line.}
\end{figure}

\subsection{Non definite photon number}
In all the above examples we have considered states with definite total photon number. These examples were addressed 
in the spirit that, in principle, intensity and polarization are independent degrees of freedom. So for simplicity we considered 
fixed total number. Nevertheless, such kind of states are difficult to generate in labs, so it would be also interesting to address 
the case of states that can be generated in practice without definite total number. This is the case of  Glauber coherent states 
and thermal states, as the most classical examples, and two-mode squeezed vacuum, as a clear  example of nonclassical light. 
For definiteness all states will be considered with the same mean total photon number $\bar{n}$.

Regarding Glauber coherent states, using SU(2) symmetry we may consider without loss of generality the product of a 
coherent state in the first mode and vacuum in the second mode so that
\begin{equation}
| C \rangle = e^{-\bar{n}/2} \sum_{n=0}^\infty \frac{\bar{n}^{n/2}}{\sqrt{n!}} |n,0 \rangle ,
\end{equation}
with $Q$ function
\begin{equation}
Q_C (\Omega ) = \frac{1}{4 \pi} e^{-\bar{n} \sin^2 \frac{\theta}{2}} \left ( 1 + \bar{n}  \cos^2 \frac{\theta}{2} \right ) .
\end{equation}
For thermal states we consider the most simple example where the second mode is also in the vacuum state:  
\begin{equation}
\rho_T  = \frac{1}{1+ \bar{n}}  \sum_{n=0}^\infty \left ( \frac{\bar{n}}{1+\bar{n}} \right )^n | n, 0 \rangle \langle n, 0 | ,
\end{equation}
with $Q$ function 
\begin{equation}
Q_T (\Omega ) = \frac{1+\bar{n}}{4 \pi} \frac{1}{\left ( 1+ \bar{n} \sin^2 \frac{\theta}{2} \right )^2} .
\end{equation}
Finally, for the squeezed vacuum state 
\begin{equation}
| S \rangle =  \frac{1}{\sqrt{1+\bar{n}/2} }\sum_{n=0}^\infty \left ( \frac{\bar{n}/2}{1+\bar{n}/2} \right )^{n/2} | n, n \rangle ,
\end{equation}
we get the $Q$ function 
\begin{equation}
Q_S (\Omega ) = \frac{\sqrt{2+\bar{n}}}{2 \pi} \frac{1}{\left ( 2+ \bar{n} \cos^2 \theta \right )^{3/2}} .
\end{equation}
With these explicit expressions it is simple to compute the ordered partial sums $S_k$ as they are plotted in Fig. 8. This shows  
that the conclusions obtained for definite total number hold also in these most realistic cases, this is that  
\begin{equation}
S \prec T \prec  C .
\end{equation} 

\begin{figure}
\centering
\includegraphics[width=7cm]{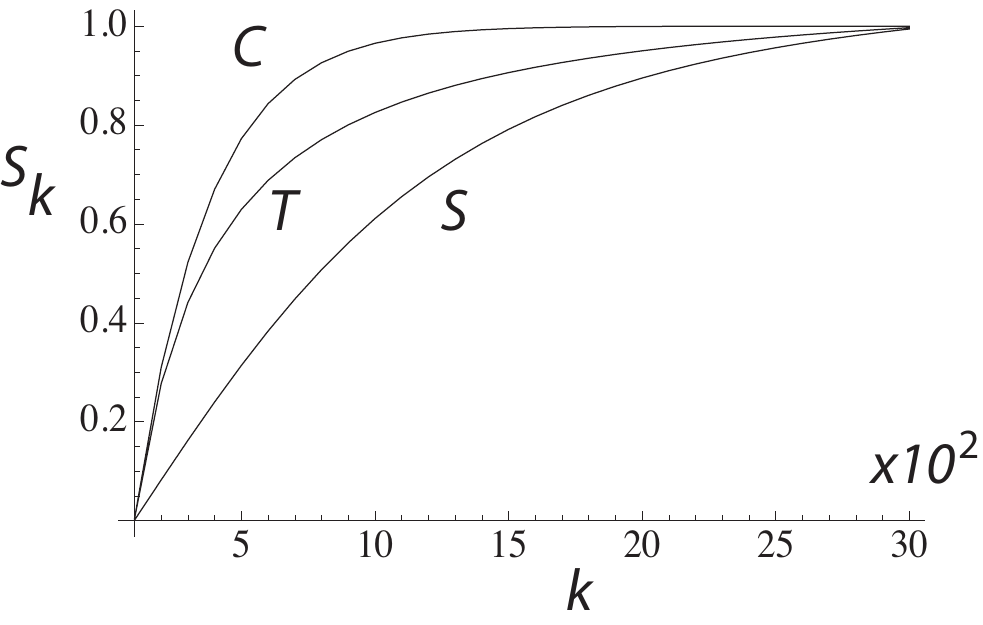}
\caption{Ordered partial sums $S_k$  as functions of $k$ for coherent $C$, thermal $T$,  and two-mode squeezed vacuum 
$S$ with the same total mean number $\bar{n} =10$.}
\end{figure}

\section{Conclusions}

We have developed the application of majorization to quantum polarization as a meta-measure of polarization fluctuations 
and degree of polarization. For fixed total number we have confirmed that the SU(2) coherent states are the most polarized 
majoring any other state. On the other hand the most nonclassical states according to the Hilbert-Schmidt distance are the 
most unpolarized among the pure states. We have shown that for odd dimension there is incompatibility between squeezed 
and $N00N$ states.  In general for states of the same class we have observed that states with larger photon numbers majorize 
states with lower numbers. Naturally, the situation is richer when comparing states of different classes and different photon 
numbers, where further cases of incomparability can be found.

\section*{ACKNOWLEDGMENTS}

G. D. gratefully acknowledges a Collaboration Grant from the spanish Ministerio de Educaci\'{o}n, Cultura y Deporte. A. L. acknowledges 
support from project FIS2012-35583 of Spanish Ministerio de Econom\'{\i}a y Competitividad and from the Comunidad 
Aut\'onoma de Madrid research consortium QUITEMAD+ S2013/ICE-2801.

\end{document}